\documentclass[twocolumn,showpacs]{revtex4}
\usepackage{psfig}
\begin{document}

\title{Charmonium photoproduction.}
\author{A. Sibirtsev$^{1,2}$, S. Krewald$^1$, and A. W. Thomas$^2$}
\affiliation{
$^1$Institut f\"ur Kernphysik, Forschungszentrum J\"ulich,
D-52425 J\"ulich \\
$^2$Special Research Centre for the Subatomic Structure of Matter (CSSM) \\
and Department of Physics and Mathematical Physics, \\
University of Adelaide, SA 5005, Australia  }

\begin{abstract}
The available data on $J/\Psi$ photoproduction are analyzed 
in terms of pomeron exchange, two gluon exchange and 
photon-gluon fusion models. Allowing the pomeron-quark 
interaction to be flavour dependent and introducing the soft and 
hard pomerons it is possible to reproduce the data at 
$\sqrt{s}{>}$10~GeV and small $|t|$. The two gluon 
exchange calculations indicate strong sensitivity 
to the gluon distribution function. The results obtained with the most
modern MRST2001 and DL PDF reproduce the forward $J/\Psi$ 
photoproduction cross section at $\sqrt{s}{>}$10~GeV.
The calculations with the photon-gluon fusion model and with
MRST2001 and DL  PDF are also in reasonable 
agreement with the data on the total $J/\Psi$ 
photoproduction cross section. We allocate the $J/\Psi$ 
photoproduction at low energies and large $|t|$ to the mechanism 
different from pomeron or two gluon exchanges.
We consider that this might be axial vector  trajectory 
exchange that couples to the  axial form factor 
of the nucleon.
\end{abstract}
\maketitle

\section{Introduction}
$J/\Psi$ production by real and virtual photons provides
an effective way to test  QCD dynamics at short distances
and to verify QCD inspired models. The pomeron exchange 
model is one of the nonperturbative QCD approaches traditionally 
applied~\cite{Donnachie1,Donnachie2,Laget1,Donnachie3,Laget2,Sibirtsev1}
to vector meson photoproduction at high energies. The
crucial point of the model is the flavour independence of the
pomeron-quark interaction, which can be studied through systematic 
analysis of the  photoproduction of light, strange, charm and 
beauty quarks. 
 
The perturbative QCD models, such  as two gluon 
exchange~\cite{Ryskin1,Brodsky,Ryskin2} and photon-gluon 
fusion~\cite{Fritzsch,Jones,Leveille,Barger,Duke,Weiler,Roberts5},
provide a direct access to the gluon distribution function. The 
$J/\Psi$ electro and photoproduction is an excellent tool to test
the gluon parton distribution function (PDF). 

$J/\Psi$ electroproduction has been systematically studied within 
these different approaches. Most modern gluon distribution functions, such 
as MRST2001~\cite{MRST2001} and CTEQ6~\cite{CTEQ6} became available
just recently through the analysis of $J/\Psi$ electroproduction
at large $Q^2$. Very recently the data on $J/\Psi$ electro production
were analyzed~\cite{Donnachie6} with the photon-gluon fusion model.

Here we present a systematic analysis of $J/\Psi$ production by real
photons. We collect old data available at low 
energies~\cite{Gittelman,Camerini,Holmes,Aubert,Frabetti} as well 
as data at high 
energies~\cite{Aid,Breitweg1,Adloff,Breitweg2} from H1 and ZEUS.
We also include very recently published results~\cite{Chekanov}.

Our aim is to investigate whether  $J/\Psi$ photoproduction can be
understood in terms of a certain model, which is also
applicable to the photoproduction of other vector mesons. It is
also a crucial line of our study whether the QCD or pQCD models are
able to explain the mechanism of $J/\Psi$ photoproduction close
to the reaction threshold, i.e. at $\sqrt{s}{<}10$~GeV.

\section{The difference between $\omega$ and $J{/}\Psi$ photoproduction.}
The $\omega$, $\phi$ and $J/\Psi$ vector mesons, which are the 
mixtures of $u{\bar u}$, $d{\bar d}$, $s{\bar s}$ and $c{\bar c}$ 
quarks states, have the same quantum numbers. For ideal mixing, 
$\theta_V$ is near 35$^o$, $\phi{=}{s\bar s}$ and $J/\Psi{=}c{\bar c}$, 
while the $\omega$ meson is built up with $u{\bar u}$ and $d{\bar d}$. 
However, the mixing is not ideal and both the $\phi$ and $J/\Psi$ 
vector mesons contain a certain fraction of light quarks.

Systematic Regge theory analysis~\cite{Sibirtsev1,Sibirtsev2} 
of $\omega$ meson photoproduction shows that
at large invariant collision energies, $\sqrt{s}{\ge}10$~GeV,
the reaction is dominated by soft pomeron exchanges, while
at $\sqrt{s}{\ge}100$~GeV additional small contribution might 
come from the hard pomeron. At low energies, $\sqrt{s}{<}10$~GeV, the
$\omega$ photoproduction is dominated by the meson exchanges.
The Regge model with contributions from $\pi$ and $f_2$ trajectories
reasonably describes the data as well as the standard meson 
exchange model with $\pi$, $\eta$ and $\sigma$ contributions.
Particularly large 
contribution from $\pi$ meson exchange to $\omega$ photoproduction
at low energies can be understood through the large 
$\omega{\to}\pi{+}\gamma$ partial decay width, which dominates 
the $\omega$ meson radiative decay mode. 

\begin{figure}[b]
\vspace*{-4mm}
\hspace*{-2mm}\psfig{file=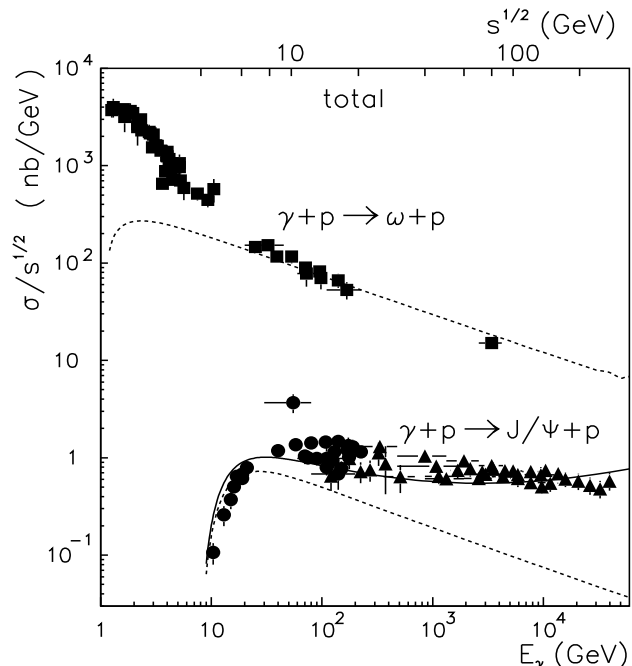,height=9cm,width=9.5cm}\vspace*{-5mm}
\caption[]{Reduced $\omega$ (squares) and $J/\Psi$ (circles and triangles) 
total photoproduction cross section as a function of photon energy. 
The $\omega$ photoproduction data are collected in Ref.~\cite{Sibirtsev1}.
The $J/\Psi$ data available before 1985 are shown by circles
and taken from Ref.\cite{Holmes}. The  $J/\Psi$ photoproduction data
at high energies are shown by triangles and are taken from
Refs.\cite{Frabetti,Aid,Breitweg1,Adloff,Chekanov}. 
The dashed lines show the calculations with soft pomeron exchange 
alone, while the solid lines are the sum of soft and hard pomeron 
contributions. The upper axis indicates the invariant collision 
energy $\sqrt{s}$.}
\label{psiju6}
\end{figure}

An absolutely different situation holds for the $J/\Psi$ meson.
Following the Regge theory it is believed that at high energies 
the $J/\Psi$ meson is produced through soft and hard pomeron exchanges,
where the last dominates at $\sqrt{s}{\ge}50$~GeV.  
It is hard to classify what mechanism is responsible for 
$J/\Psi$ photoproduction at low energies, below $\sqrt{s}{=}10$~GeV.
The $J/\Psi$ mass $m_J{\simeq}3.097$~GeV is below  the $D{\bar D}$ threshold
${\simeq}3.739$~GeV and OZI~\cite{Okubo,Zweig,Iizuka,Sibirtsev4} allowed 
$J{/}\Psi{\to}D{+}{\bar D}$ decay cannot not be 
realized in vacuum. Instead there are large amount of
OZI suppressed decay channels involving stable hadrons, 
hadronic resonances and radiative decays~\cite{PDG}. 
A large fraction of radiative $J/\Psi$ decay proceeds through 
the production of multi mesonic resonant and non resonant states. 
The $J/\Psi{\to}\pi{+}\gamma$ and $J/\Psi{\to}\eta{+}\gamma$ both 
constitute a negligible part of the total radiative decay modes,
thus the  single meson exchanges contribution to $J/\Psi$
photoproduction are expected to be very small.

However, it might be that the mesonic or multi mesonic exchange 
contribution to $\gamma{+}N{\to}J{/}\Psi{+}N$ is  
negligible and the soft pomeron exchange dominates already 
starting from the reaction threshold.

The difference between $\omega$ and $J{/}\Psi$ total 
photoproduction cross section is illustrated by Fig.\ref{psiju6}, where 
the reduced total photoproduction cross section $\sigma_f{/}\sqrt{s}$ is 
shown as a function of photon energy, $E_\gamma$. The upper axis of
Fig.\ref{psiju6} indicates the  invariant collision energy $\sqrt{s}$, 
where $s{=}m_N^2{+}2m_NE_\gamma$ with $m_N$ being the nucleon mass. 

The $\omega$ photoproduction data are shown by squares and were 
collected in Ref.~\cite{Sibirtsev1}. The  circles show the $J{/}\Psi$ 
data taken from  Ref.~\cite{Holmes}, where the results available until 
1985 were reviewed. The part of the $J{/}\Psi$ photoproduction 
data~\cite{Holmes} were obtained with the nuclear target and 
therefore should be considered as an average between the proton 
and neutron.
The $J{/}\Psi$ photoproduction threshold 
is $E_\gamma{\simeq}8.15$~GeV. The Cornell measurement~\cite{Gittelman} 
is the only result available now for $J{/}\Psi$ photoproduction close 
to the reaction threshold.
We note that the  $J{/}\Psi$ photoproduction 
data measured  at $9.0{\le}E_\gamma{\le}11.8$~GeV at Cornell were 
incorrectly reviewed in Ref.~\cite{Holmes}. The original experimental
numbers for the differential $J{/}\Psi$ photoproduction cross section
were published~\cite{Gittelman} in the form 
$d\sigma{/}dt{=}(1.01{\pm}0.20)$$\exp[(1.25{\pm}0.20)t]$ 
nb/GeV$^2$. The total photoproduction cross section is therefore  
$\sigma{=}0.48{\pm}0.12$~nb, taking into account the minimal
four momentum transfer squared. 

The triangles in Fig.\ref{psiju6} show the $J{/}\Psi$ photoproduction 
data, which  appeared after 1985 and are published in 
Refs.\cite{Frabetti,Aid,Breitweg1,Adloff,Breitweg2,Chekanov}. 
It is clear that the energy dependence of  
$\gamma{+}p{\to}\omega{+}p$ and $\gamma{+}N{\to}J{/}\Psi{+}N$ is 
different at all photon energies. It is known that this difference 
was discovered after the H1 and ZEUS 
measurements~\cite{Aid,Breitweg1,Adloff,Breitweg2,Chekanov} 
at $\sqrt{s}{\ge}50$~GeV.

\begin{figure}[t]
\phantom{aa}
\hspace*{-4mm}\psfig{file=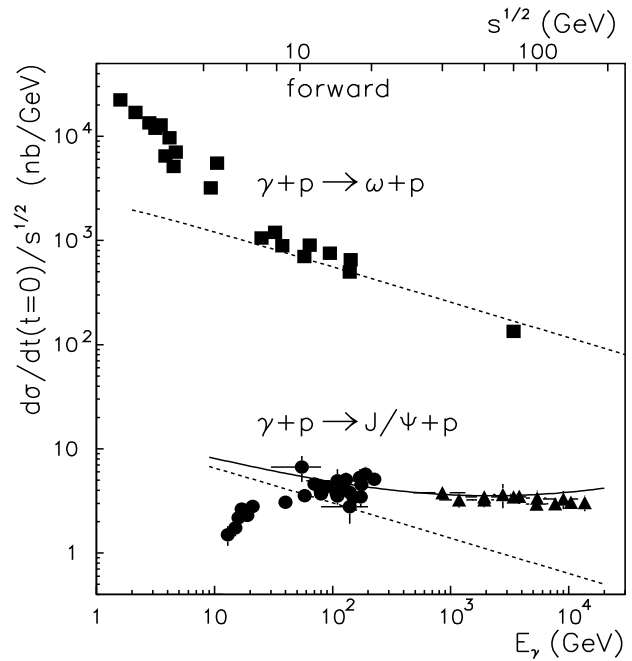,height=9cm,width=9.5cm}\vspace*{-5mm}
\caption[]{Reduced $\omega$ (squares) and $J/\Psi$ (circles and
triangles) forward photoproduction cross sections as a function of 
photon energy. The $\omega$ photoproduction data are collected in
Ref.\cite{Sibirtsev1}. The circles show $J/\Psi$ data from 
Ref.\cite{Holmes}, while triangles show $J/\Psi$ photoproduction 
data from Refs.\cite{Adloff,Chekanov}. The dashed 
lines show the calculations with soft pomeron exchange alone, while the
solid lines are the sum of soft and hard pomeron contributions.}
\label{psiju7}
\end{figure}

Fig.\ref{psiju7} show the $\omega$ and $J{/}\Psi$ reduced differential 
photoproduction cross section, $d\sigma{/}dt{/}\sqrt{s}$, extrapolated to
zero four momentum transfer squared $t{=}0$ as a function of 
photon energy $E_\gamma$ and invariant collision energy $\sqrt{s}$.
The energy dependence of the forward photoproduction cross sections 
again indicates a strong difference between the $\omega$ and 
$J{/}\Psi$ data. Our meaning of 
the forward photoproduction cross section is an extrapolation of 
$d\sigma{/}dt$ to $t{=}0$.

\begin{figure}[t]
\vspace*{2mm}
\hspace*{-4mm}\psfig{file=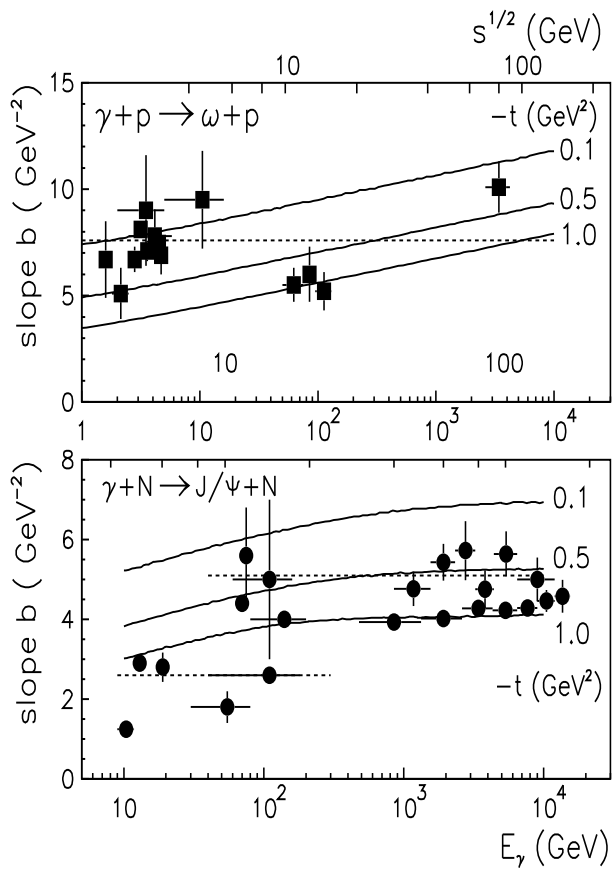,height=11.8cm,width=9.5cm}\vspace*{-5mm}
\caption[]{The slope, $b$, of the $t$ dependence of the differential $\omega$ 
(squares) and $J/\Psi$ (circles) photoproduction cross sections as a 
function of photon energy, $E_\gamma$, and invariant collision energy, 
$\sqrt{s}$. The data are from Refs.\cite{Sibirtsev1,Holmes,Adloff,Chekanov}.
The solid lines show the calculations with soft pomeron exchange 
alone for $\omega$ photoproduction and with the sum of soft and hard 
pomeron contributions for $J/\Psi$ photoproduction. The calculations 
show the energy dependence of the local slope at the indicated
four momentum transfer squared, $t$. The dashed lines show the fits using
a constant value. The $J/\Psi$ photoproduction data were fitted 
separately at photon energies below and above 100~GeV.}
\label{psiju2a}
\end{figure}

Another value of interest is the exponential slope 
$b$ of $t$ dependence of the differential photoproduction 
cross section, which is defined as
\begin{equation}
\frac{d\sigma}{dt} = \left. \frac{d\sigma}{dt} 
\right|_{t=0}\!\!\!\!\times\exp(bt),
\end{equation}
and is shown in Fig.\ref{psiju2a}. The $\omega$ photoproduction data 
can be well fitted with a constant slope $b{=}7.6{\pm}0.1$~GeV$^{-2}$
over a wide range of available photon energies. In contrast, the $J/\Psi$ 
photoproduction data show a strong energy dependence of the exponential 
slope. At $E_\gamma{<}100$~GeV the data can be well reproduced
with a constant slope $b{=}2.6{\pm}0.4$~GeV$^{-2}$, while at
high energies $b{=}5.1{\pm}0.2$~GeV$^{-2}$.

Finally, the available experimental results on $\omega$ and $J/\Psi$
meson photoproduction might indicate  a rather different mechanism
for their production. Although, the quantum numbers of $\omega$ and 
$J/\Psi$ are the same, they have different quark structures.
The difference between the $\omega$ and $J/\Psi$ photoproduction data 
can be attributed to the difference between the properties
of light and charm quarks, such as their interaction through the
gluon exchange, confinement and fusion into  real hadrons and 
photon quark-antiquark fluctuation. Therefore, currently available 
theoretical models for $\omega$ and $J/\Psi$ meson photoproduction
are inspired by QCD. 

In the next section we review some of the models that recently were 
successfully applied to the $J/\Psi$ electroproduction off the nucleon.
We will compare these models with data on $J/\Psi$ photoproduction
by real photons. Furthermore, we consider all available
data on $J/\Psi$ photoproduction in order to make our study in
a more systematic way. 

\section{Pomeron exchange model.}
The basis of the model is the factorization of the exclusive vector 
meson photoproduction amplitude in terms of the product
of $\gamma{\to}q{+}{\bar q}$ fluctuation, the scattering of the 
$q{\bar q}$ system by the proton and finally the $q{\bar q}$
hadronization into a vector meson. The interaction
between the $q{\bar q}$ state confined in the vector meson and the
nucleon is taken to be given by  pomeron exchange. The pomeron 
propagator $G(s,t)$ is parameterized as
\begin{equation}
G(s,t) = i \left( \frac{s}{s_0}\right)^
{\alpha (t)-1}\!\!\!\exp(-\frac{i\pi}{2}[\alpha(t)-1]),
\label{base}
\end{equation}
and therefore the pomeron trajectory  $\alpha(t)$ drives
the $\sqrt{s}$ energy dependence of the reaction amplitude. 
The evolution of $t$ dependence with energy is 
controlled by pomeron exchange trajectory as well.

The interaction between the pomeron and quark is considered to be
a flavour independent. The flavour independence of the 
quark-pomeron interaction might provide a unique way to construct
a unified model for vector meson photoproduction.

\subsection{Soft pomeron.}
\begin{figure}[b]
\vspace*{-15mm}
\hspace*{-9mm}\psfig{file=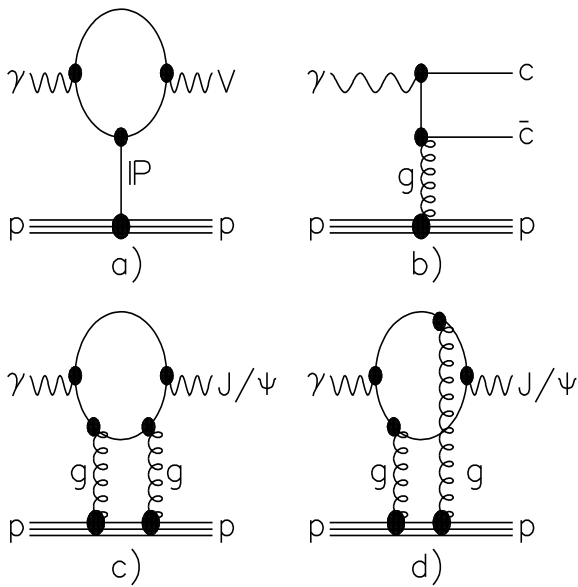,height=10cm,width=9.5cm}\vspace*{-11mm}
\caption[]{The diagrams for pomeron exchange (a), two gluon exchange
c,d) and photon-gluon fusion model b).}
\label{psiju14}
\end{figure}

The pomeron  exchange amplitude ${\cal T}$ for vector meson 
photoproduction is depicted in Fig.\ref{psiju14}a) and is explicitly 
given~\cite{Donnachie1,Donnachie2,Laget1,Donnachie3,Laget2} 
in the form
\begin{eqnarray}
{\cal T}{=}3 i F_1(t)\, \frac{8  \sqrt{6}\, m_q e_q f_V 
\beta_q^2}
{4m_q^2-t} (\varepsilon{\cdot}\varepsilon_V) \, \, 
\left( \frac{s}{s_0}\right)^{\alpha (t)-1} \nonumber \\ 
\times 
\exp(-\frac{i\pi}{2}[\alpha(t)-1])\,
\frac{\mu_q^2}{2\mu_q^2+4m_q^2-t},
\label{spom}
\end{eqnarray}
where $e_q$ and $m_q$ are the charge and mass of the 
quark, $\varepsilon$ and $\varepsilon_V$ are the 
polarization vectors of the photon and vector meson, respectively 
and $f_V$ is the meson decay constant given by $V{\to}e^+e^-$ 
radiative decay width $\Gamma_{e^+e^-}$ as
\begin{equation}
\Gamma_{e^+e^-}=\frac{8\pi \alpha^2 e_q^2}{3m_V}f_V^2,
\label{meson}
\end{equation}
where $\alpha$ is the electromagnetic coupling constant and 
$m_V$ is the vector meson mass.

Furthermore, $t$ is the squared four momentum transfer
and $F_1(t)$ is  the proton isoscalar EM form factor given as
\begin{equation}
F_1(t)=\frac{4m_p^2-2.8t}{4m_p^2-t} \frac{1}{(1-t/t_0)^2},
\label{emp}
\end{equation}
with $m_p$ being the proton mass and $t_0$=0.71~GeV$^2$. The
$F_1(t)$ was introduced assuming the pomeron resembles an 
isoscalar photon.

The pomeron trajectory $\alpha (t)$ for soft pomeron 
exchange is given as
\begin{equation}
\alpha (t)=1.08+\alpha^\prime t,
\label{tra1}
\end{equation}
with $\alpha^\prime$=0.25~GeV$^{-2}$.
The constant $s_0$ is not well defined theoretically and 
can be taken as $s_0{=}1/\alpha^\prime$ following the
dual model prescription~\cite{Veneziano}.

The free model parameter, $\beta_q$, determines 
the strength of the effective pomeron coupling to the quark,
while $\mu_q$ accounts for the possibility that the coupling to 
an off-shell quark is not pointlike but dressed with the form factor 
given by the last term of Eq.(\ref{spom}). Both $\beta_q$ and, in 
principle, $\mu_q$ might depend on the flavour of the quark and can be
fixed by experimental data.

The phase of the soft pomeron exchange amplitude was not written explicitly 
in Refs.~\cite{Donnachie1,Donnachie2,Laget1,Donnachie3,Laget2} 
since no interference with other exchanges was considered.

The amplitude is normalized so that $d\sigma{/}dt{=}\alpha|{\cal T}|^2$.
Finally, the differential $\gamma{+}p{\to}V{+}p$ cross section 
due to  soft pomeron exchange is given for a real photon as, 
\begin{eqnarray}
\frac{d\sigma}{dt}=\frac{81\, m_V^3 \, \beta_q^4 \,\, \mu_q^4 \,\,
\Gamma_{e^+e^-}}
{\pi \alpha} \, \left( \frac{s}{s_0}
\right)^{2\alpha_{P_1}(t)-2}\nonumber \\ 
\times \frac{F_1^2(t)}{(m_V^2-t)^2\,\,
(2\mu_q^2+m_V^2-t)^2},
\label{pom1}
\end{eqnarray}
where it was assumed that $m_V{=}2m_q$.

The coupling constant  $\beta_q$ and  the form factor $\mu_q$ 
were fitted~\cite{Sibirtsev2} through the systematical analysis 
of $\omega$ meson photoproduction as 
\begin{equation}
\beta_q=2.35~\mbox{GeV}^{-1}\!\!, \hspace{7mm} \mu_q^2=1.1~\mbox{GeV}.
\label{par}
\end{equation}
It was found~\cite{Sibirtsev2} that the 
parameter $\beta_q$ evaluated from $\omega$ photoproduction was
different from $\beta_q$=2.0~GeV$^{-1}$ needed in the 
description~\cite{Donnachie4,Donnachie5} of $\rho$ photoproduction data. 
This discrepancy could not be explained in a satisfactory way, since 
once the parameters of the pomeron exchange are fixed by $\rho$ meson 
photoproduction, they should be applicable to the other vector mesons 
as well. Moreover, at this stage  we still do not discuss the flavour 
dependence of the quark-pomeron interaction, because we are dealing 
with $\rho$ and $\omega$ photoproduction, i.e. light $q\bar{q}$ states.  

\begin{figure}[t]
\vspace*{-1mm}
\hspace*{-4mm}\psfig{file=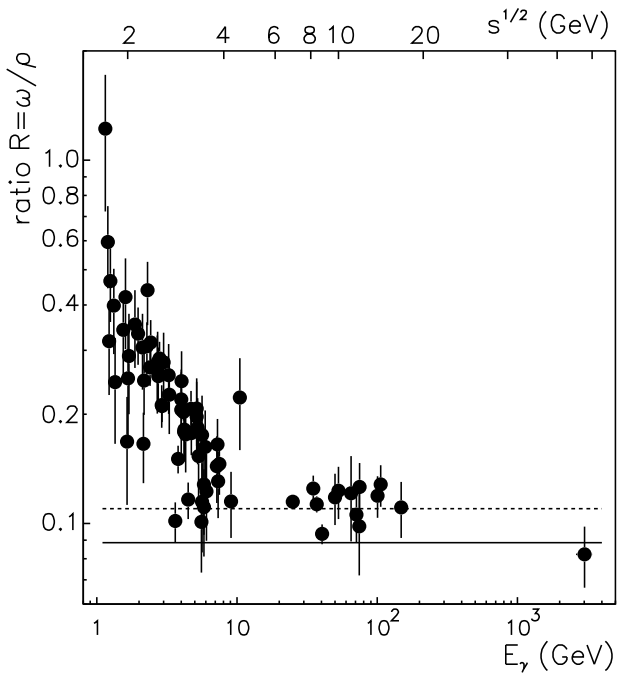,height=9.5cm,width=9.5cm}\vspace*{-6mm}
\caption[]{The ratio of $\gamma{+}p{\to}\omega{+}p$ to
$\gamma{+}p{\to}\rho{+}p$ cross sections as a 
function of photon energy, $E_\gamma$, and invariant collision energy, 
$\sqrt{s}$. The solid line show the ratio of the $\omega{\to}e^+e^-$
and $\rho{\to}e^+e^-$ decay widths. The dashed line shows the fit
to data with a constant value of $R{=}0.115{\pm}0.003$.}
\label{reg12a}
\end{figure}

An ideal illustration of this finding is
given by the ratio of the $\gamma{+}p{\to}\omega{+}p$ and 
$\gamma{+}p{\to}\rho{+}p$ cross sections, shown in Fig.~\ref{reg12a}
as a function of photon energy $E_\gamma$ and invariant collision energy 
$\sqrt{s}$. Following Eq.(\ref{pom1}) the ratio of $\omega$ and
$\rho$ meson photoproduction is given by the ratio  of 
$\omega{\to}e^+e^-$ and $\rho{\to}e^+e^-$ decay widths
\begin{equation}
R\left(\frac{\omega}{\rho}\right) {=}\frac {\Gamma_{\omega{\to}e^+e^-}}
{\Gamma_{\rho{\to}e^+e^-}}
{=}\frac{0.60{\pm}0.02~\mbox{keV}}{6.77{\pm}0.32~\mbox{keV}}
=0.088{\pm}0.005,
\end{equation}
which is shown by the solid line in Fig.~\ref{reg12a} and considerably
underestimates the experimental results. 

The dashed line in Fig.~\ref{reg12a} shows the fit to the data
by a constant value $R{=}0.115{\pm}0.003$, which is close to the 
$SU(3)$ estimates for the $\omega$ and $\rho$ coupling to the photon.
Finally the difference between the Regge model and the data accounts 
for a factor of $\simeq$1.3. It is already clear that both $\rho$ 
and $\omega$ photoproduction data could not be reproduced 
by the soft pomeron exchange simultaneously with the same set of 
parameters $\beta_q$ and $\mu_q$. 

Now, the dashed line in Fig.~\ref{psiju6}
shows the contribution from soft pomeron exchange to the
total reduced $\omega$ meson photoproduction cross section. 
The calculations were done with the parameters
given by Eq.(\ref{par}). The comparison to differential 
$\gamma{+}p{\to}\omega{+}p$ cross section at different photon energies
is given in Ref.~\cite{Sibirtsev2}. The contribution from soft
pomeron exchange to forward $\omega$ meson photoproduction is shown in
Fig.~\ref{psiju7} by the dashed line.

Obviously, the soft pomeron exchange
alone will reproduce the $\omega$ photoproduction data at
$\sqrt{s}{\ge}6$~GeV. At lower energies the dominant contribution
to the $\gamma{+}p{\to}\omega{+}p$ reaction comes from meson 
exchanges.

The $t$ dependence of the photoproduction cross section
arising from the soft pomeron exchange  is given by
the proton EM form factor squared $F_1^2(t)$, the quark-pomeron
form factor squared and by the pomeron
exchange trajectory $(s{/}s_0)^{2\alpha(t){-}2}$. The exponential 
slope, $b$, of the $t$ dependence can be evaluated as
\begin{equation}
b=\frac{d}{dt} \ln \left[ \frac{d\sigma}{dt} \right],
\end{equation}
from the differential photoproduction cross section given by
Eq.(\ref{pom1}).

The slope due to the proton isoscalar EM form factor is
\begin{equation}
b_1= -\frac{5.6}{4m_p^2-2.8t} + \frac{2}{4m_p^2-t}
+\frac{4}{t_0-t},
\label{slopeb1}
\end{equation}  
and since it depends on $t$ it can be considered as a local slope. 
At $t{=}0$ the slope from proton EM form factor is $\simeq$4.6.
The slope due to the quark-pomeron form factor squared is given as
\begin{equation}
b_2=\frac{2}{m_V^2-t}+\frac{2}{2\mu_q^2+m_V^2-t},
\end{equation}
and depends on the cutoff parameter $\mu_q$ and mass of the produced 
vector meson $m_V$. For $\omega$ meson photoproduction at $t$=0
the slope associated with the quark-pomeron form factor is $\simeq$4.0.
The energy dependence of the slope comes from pomeron trajectory
and is given for soft pomeron exchange as
\begin{equation}
b_3 = 0.5 \ln (0.25 s).
\end{equation}

Finally, the total exponential slope, $b$, of the $t$ dependence for
$\omega$ meson photoproduction at $t$=0 has a minimum value of
$b$=8.6~GeV$^{-2}$ and increases logarithmically with energy. This 
result is in excellent agreement with the data shown in 
Fig.\ref{psiju2a}. Here we also show the local slope $b$ at 
different squared four momentum transfers $t$. It is also clear 
that the appearance of the proton EM form factor in the reaction 
amplitude dictates quite a large value of the minimal 
available $b$ at $t$=0.

The contribution from soft pomeron exchange to $J/\Psi$ photoproduction
is given by Eq.(\ref{pom1}). Taking the parameters $\beta_q$ and
$\mu_q$ as determined from the data on $\omega$ meson photoproduction
one can easy estimate the ratio of forward $J/\Psi$ and  $\omega$
photoproduction cross section as
\begin{equation}
R\left( \frac{J/\Psi}{\omega}\right){=}
\frac {\Gamma_{J{\to}e^+e^-}}{\Gamma_{\omega{\to}e^+e^-}}
\frac{m_\omega (m_\omega^2+2\mu_q^2)}{m_J (m_J^2+2\mu_q^2)}
{=}0.13,
\end{equation}
with $\Gamma_{J{\to}e^+e^-}$=5.26~keV. The experimental results
shown in Fig.\ref{psiju7} indicate that this  ratio 
is of order 0.005 at
$\sqrt{s}{\simeq}$10~GeV. Therefore the soft pomeron exchange with 
fixed parameters for the quark-pomeron interaction substantially 
overestimates the data on $J/\Psi$ photoproduction. However, since it was
already realized that a simultaneous description of both $\rho$ and 
$\omega$ photoproduction data requires a different coupling constant
$\beta_q$, one might proceed in a similar way and readjust $\beta_q$
to the $J/\Psi$ photoproduction data. 

Considering the flavour dependence of the interaction between the 
pomeron and quark one might distinguish the light quark coupling
$\beta_u$=$\beta_d$=2.35~GeV$^{-1}$, which is taken to be the same 
as evaluated from $\omega$ meson photoproduction, and the 
pomeron coupling to charm quark $\beta_c$. In that case the 
$b_q^2$ term in the amplitude given by Eq.(\ref{spom}) should 
be replaced by product $b_ub_c$. In addition the differential 
cross section of Eq.(\ref{pom1}) must also be corrected. 

We found that the choice of $\beta_c$=0.45~GeV$^{-1}$ provides 
a reasonable description of the data around $\sqrt{s}{\simeq}$10~GeV.
The dashed lines in Figs.\ref{psiju6},\ref{psiju7} show our
calculation of the  soft pomeron contribution to the reduced
total and forward $J/\Psi$ photoproduction cross section.

However, resolving the problem with the reduction of the soft
pomeron contribution to $J/\Psi$ photoproduction leaves one 
far short of a description of the $J/\Psi$ meson data 
collected in Figs.\ref{psiju6},\ref{psiju7},\ref{psiju2a}. 

First, although  the soft pomeron seems to reproduce
the total $J/\Psi$ photoproduction cross section starting 
just from the reaction threshold, the model absolutely fails
to describe the forward cross section and slope $b$ of the $t$ 
dependence at low energies. Recall, that the minimal value of the
slope is given by the proton scalar EM form factor and this is
already larger than the experimental results at low energies.

Second, the energy dependence of the total and forward $J/\Psi$ 
photoproduction  cross section and  the slope $b$ at high energies 
cannot be reproduced by soft pomeron exchange. 
   
\subsection{Hard pomeron.}
There are different attempts to reproduce the energy dependence
of the total  $J/\Psi$ photoproduction  cross section without  
introducing a new pomeron trajectory. The strategy is to
readjust the $t$ dependence in such a way that it becomes flatter,
providing the correct total cross section after integration over 
the large $t$ region
accessible at high energies. One of the more natural ways is
to introduce a trajectory that is  nonlinear in $t$, instead of
that given by Eq.(\ref{tra1}). Obviously the $J/\Psi$ 
photoproduction data at $t$=0, shown in Fig.\ref{psiju7}, 
do not support such solution.

\begin{figure}[b]
\vspace*{-8mm}
\hspace*{-5mm}\psfig{file=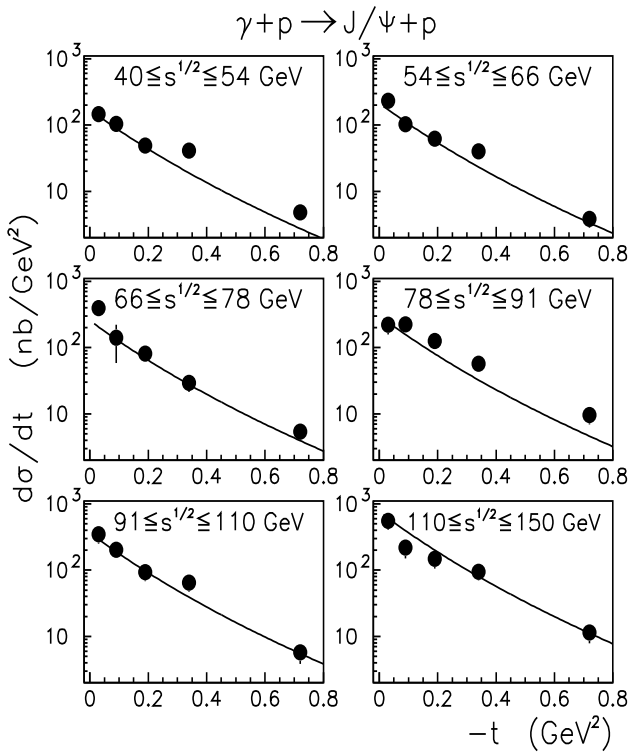,height=10.9cm,width=9.5cm}\vspace*{-7mm}
\caption[]{The $\gamma{+}p{\to}J/\Psi{+}p$ differential cross
section as a function of four momentum transfer squared, $t$,
measured at different invariant collision energies $\sqrt{s}$
by H1 Collaboration~\cite{Adloff}. The solid lines show the 
calculations including both soft and hard pomeron exchanges.}
\label{psiju4}
\end{figure}

Instead, the forward $J/\Psi$  photoproduction cross section 
requires that at $t$=0 the cross section is proportional
to $\sqrt{s}$, because the reduced cross section in Fig.\ref{psiju7} 
is almost independent of energy for $\sqrt{s}{>}$10~GeV. From
Eq.(\ref{base}) one can show that the trajectory necessary to
reproduce the data has an intercept $\alpha(t{=}0){\simeq}$1.25.
Moreover, since the energy dependence of the slope, $b$, is
proportional to $\alpha^\prime\ln(s)$, the data shown in
Fig.\ref{psiju2a} support the value of
$\alpha^\prime{\simeq}$0~GeV$^{-2}$.

The hard pomeron trajectory was introduced in Ref.~\cite{Donnachie3} as
\begin{equation}
\alpha (t)=1.44 +  0.1 t.
\label{tra2}
\end{equation}
The calculations with both hard and soft pomeron exchanges,
including their interference, are shown by the solid lines in 
Figs.\ref{psiju6},\ref{psiju7},\ref{psiju2a}. The coupling
constant between the hard pomeron and charm quark was fitted
as ${\tilde \beta_c}$=0.05~GeV$^{-1}$. The calculations 
reproduce quite well the data on total and forward $J/\Psi$ 
photoproduction cross section and slope $b$ of $t$ dependence at
$\sqrt{s}{\ge}$10~GeV. Note that for both soft and hard
pomeron exchanges the partial slope $b_1$, coming from the
form factor in the upper pomeron-quark vertices, is
$b_1{\simeq}$0.37, because of the large $J/\Psi$ mass appearing in
Eq.(\ref{slopeb1}). It is clear that the hard pomeron dominates 
at high energies.

Figs.\ref{psiju4} shows the differential cross section for 
$J/\Psi$ photoproduction at high energies. The data available at
small $t$ can be described well by the model.

While the introduction of a hard pomeron exchange allows one to 
resolve inconsistencies at high energies, the discrepancies 
between the model and the data at $\sqrt{s}{<}$10~GeV still
remain. 

\begin{figure}[t]
\vspace*{1mm}
\hspace*{-5mm}\psfig{file=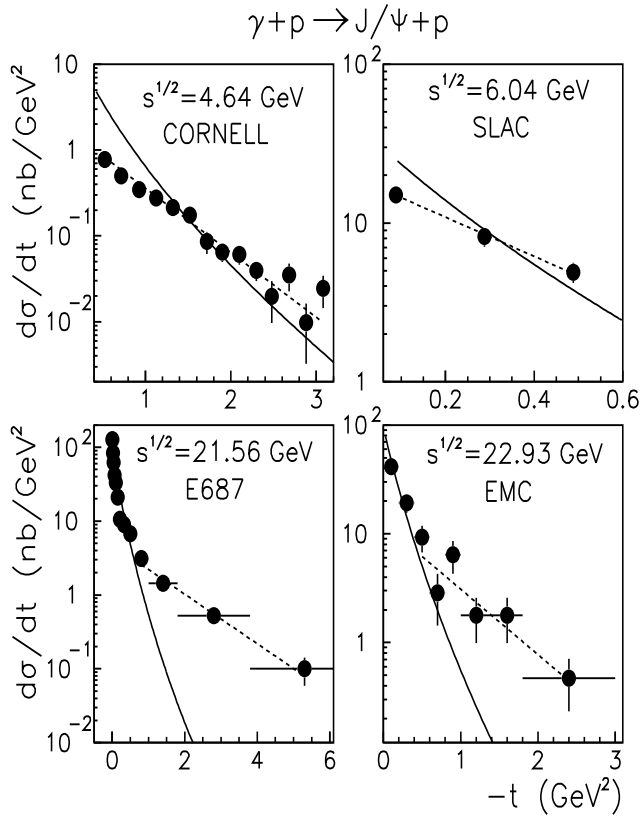,height=11.5cm,width=9.5cm}\vspace*{-7mm}
\caption[]{The $\gamma{+}p{\to}J/\Psi{+}p$ differential cross
section as a function of four momentum transfer squared $t$
measured at different invariant collision energies $\sqrt{s}$.
The data are from Cornell~\cite{Gittelman}, E687~\cite{Frabetti},
EMC~\cite{Aubert} and SLAC Collaboration~\cite{Camerini}.
The solid lines show the calculations including both soft and hard pomeron 
exchanges. The dashed lines indicate the fit to the soft part of the
spectra.}
\label{psiju8}
\end{figure}

The most crucial data are shown in Fig.\ref{psiju8},
where we collect available data on differential
$\gamma{+}N{\to}J/\Psi{+}N$  cross section at relatively small 
energies, $\sqrt{s}{<}$23~GeV. The solid lines
show the calculations with soft and hard pomeron contributions.
The dashed lines show the fit to the soft part of the differential
spectra. 

The comparison between the differential spectra and calculations
at $\sqrt{s}{<}$10~GeV illustrates that both slope and extrapolated
cross section at $t$=0 could not be reproduced by the model. The
rough agreement between the integrated theoretical and experimental
cross sections shown in Fig.\ref{psiju6} should be considered as 
an accident. The pomeron exchange model is far from  describing 
the data at $\sqrt{s}{<}$10~GeV.

The comparison at $\sqrt{s}{\simeq}$20~GeV shows that pomeron exchange 
allows one to describe the steep dependence of the differential 
$\gamma{+}N{\to}J/\Psi{+}N$ cross section at small $t$. At large
four momentum transfer squared there are  additional contributions
resulting in a soft component of the spectrum. It is worthwhile to
note the slope of the soft component observed 
at $\sqrt{s}{\simeq}$20~GeV is close to that measured at
lower energies and is around 0.7$\div$2.8~GeV$^{-2}$. 

\begin{figure}[b]
\vspace*{-5mm}
\hspace*{-3mm}\psfig{file=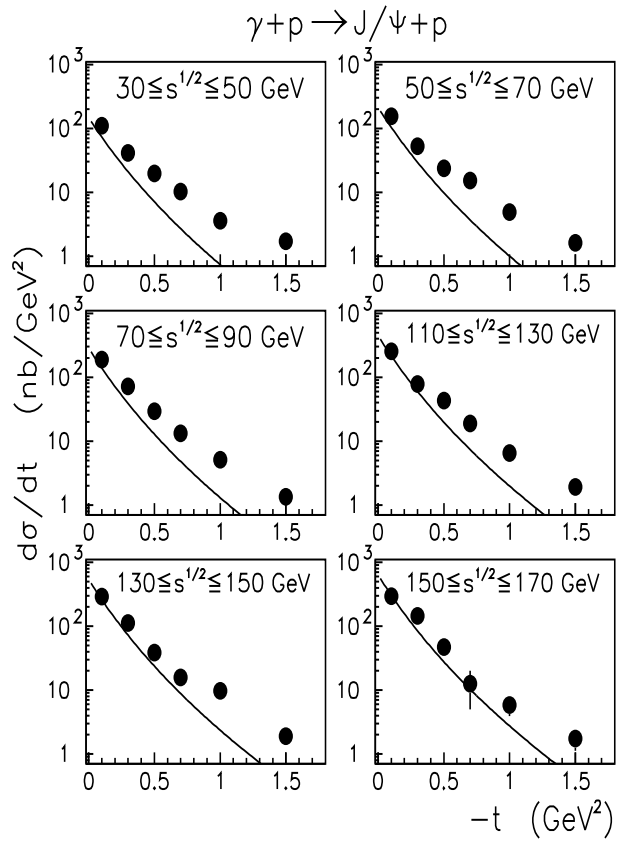,height=11.9cm,width=9.5cm}\vspace*{-7mm}
\caption[]{The $\gamma{+}p{\to}J/\Psi{+}p$ differential cross
section as a function of four momentum transfer squared $t$
measured at different invariant collision energies $\sqrt{s}$
by ZEUS Collaboration~\cite{Chekanov}. The solid lines show the 
calculations including both soft and hard pomeron exchanges.}
\label{psiju4a}
\end{figure}

Fig.\ref{psiju4a} shows recent data on differential
$\gamma{+}N{\to}J/\Psi{+}N$  cross section collected by ZEUS 
Collaboration~\cite{Chekanov} at $30{\le}\sqrt{s}{\le}170$~GeV.
The solid lines show the calculations with soft and hard pomeron 
contributions, which well describe the data at low $|t|$.
At large four momentum transfer squared we detect a soft 
component of the spectrum~\cite{Kochelev1,Kochelev2,Kochelev3}. 

\subsection{Short comment.}
It is clear that assuming the appropriate flavour dependence of the 
interaction between pomeron and quark the Regge model allows
an excellent description of the high energy data on vector meson 
photoproduction. 

There are attempts to recover flavour
independence by accounting for the mass-dependent corrections,
such as Fermi motion and quark off-shellness in the $q{\bar q}$ loop.
Indeed the vector meson coupling to $q{\bar q}$ introduced through
Eq.(\ref{meson}) was obtained within an on-shell approximation,
which is not correct in the case of diffractive photoproduction. 
However, it was found that calculations~\cite{Pichowsky,Royen,Donnachie8}
with realistic vector meson wave functions do not allow one to describe 
simultaneously the $\rho$ and $J/\Psi$ meson electroproduction,
similar to what we have found in the present study of 
$\omega$ and $J/\Psi$ photoproduction by real photons. It was 
proposed to use a Gaussian wave function instead and to
regulate its mean value individually for each vector meson. 
It is not clear whether such degree of freedom is better than 
the individual choice of coupling constant $\beta_q$.

\section{Two gluon exchange model.}
The exclusive vector meson photoproduction amplitude 
is again factorized in terms 
of the $\gamma{\to}q{+}{\bar q}$ fluctuation, the scattering of the 
$q{\bar q}$ system by the proton and finally the $q{\bar q}$
hadronization into the vector meson. The  $q{\bar q}$ system
interacts with a nucleon through  two gluon exchange. 

In lowest order perturbative QCD the photoproduction amplitude
is given by the sum of two diagrams depicted in Fig.\ref{psiju14}b,c)
and can be written as~\cite{Ryskin1,Brodsky,Ryskin2}
\begin{eqnarray}
{\cal T} = \frac{i\, 2\,\sqrt{2}\, \pi^2}{3}\,  
m_q \,\alpha_s \, \, e_q f_V\, F_{2g}(t)
\nonumber \\
\times \int \!\!dl^2 D_g^2(l) \, [D_+(l){-}D_-(l)] \, G(l),
\end{eqnarray}
where  the integration is performed over the gluon 
transverse momentum, $e_q$ and $m_q$ are the charge and mass 
of the quark, respectively, while $\alpha_s$ is the QCD 
coupling constant~\cite{PDG}. The 
meson decay constant is given by Eq.(\ref{meson})
The gluon propagator $D_g(l)$ is taken as $1/l^2$, the
$D_-(l)$ is the propagator of the off-shell quark
in the diagram where each gluon couples to a different quark
of the vector meson and  is given as
\begin{equation}
D_-(l) =(-2m_q^2-2l^2)^{-1}.
\end{equation}
When the two gluons couple to the same quark the off-shell quark
propagator is
\begin{equation}
D_+(l) =(-2m_q^2)^{-1}.
\end{equation}

Furthermore, $F_{2g}(t)$ accounts for the $t$ dependence of the 
amplitude given by a two gluon correlation in the proton. The form
factor $F_{2g}(t)$ is not defined by the model and,
as proposed in Ref.\cite{Ryskin1}  can be taken to be the 
proton isoscalar EM form factor, $F_1(t)$, given by Eq.(\ref{emp}).

Function $G(l)$ defines the probability of catching the two gluons with
momenta $l$ from the  proton and it is related to the conventional 
gluon distribution function $g(x)$ as~\cite{Ryskin1,Brodsky}
\begin{equation}
{x} g(x,Q^2) = \int\limits^{Q^2} \! dl^2 \,  \frac{G(l)}{l^2}, 
\end{equation}

Finally the two gluon exchange amplitude becomes
\begin{eqnarray}
{\cal T} = \frac{i\, \sqrt{2}\, \pi^2}{3}\,  
m_q \,\alpha_s \, \, e_q f_V\, F_{2g}(t)
\nonumber \\
\times\left[ \frac{xg(x,Q^2_0)}{m_q^4}+
\int\limits_{Q^2_0}^{+\infty} \!\!\frac{dl^2}{m_q^2(m_q^2+l^2)} \,
\frac{\partial x g(x,l^2)}{\partial l^2} \right]
\end{eqnarray}

The amplitude is normalized so that $d\sigma{/}dt{=}\alpha|{\cal T}|^2$
and in lowest order the $J/\Psi$ photoproduction cross section 
is given as  
\begin{equation}
\frac{d\sigma}{dt}{=}\frac{\pi^3 \Gamma_{e^+e^-}\alpha_s}
{6\alpha m_q^5}\left[x g(x,Q_0^2\right]^2,
\label{twog}
\end{equation}
where $x$=$m_J^2/s$ and $Q_0{\simeq}m_V$. The 
cross section depends on the gluon distribution function 
squared and the energy dependence of $d\sigma{/}dt$ is
directly given by the $x$ dependence of $xg(x)$.

\begin{figure}[t]
\vspace*{4mm}
\hspace*{-5mm}\psfig{file=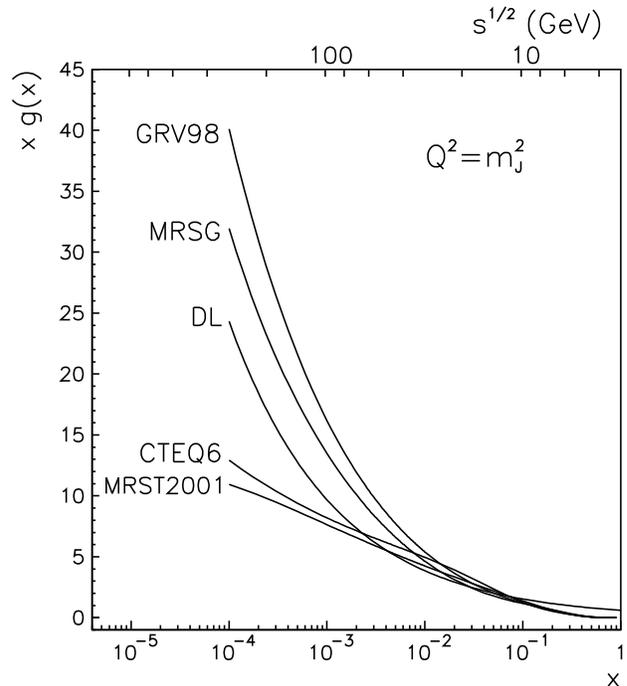,height=9.5cm,width=9.5cm}\vspace*{-7mm}
\caption[]{The gluon distribution functions as a function of $x$
at $Q^2{=}m_J^2$. Upper axis shows the relevant invariant collision
energy given as $\sqrt{s}{=}m_J{/}\sqrt{x}$. The functions are
taken from Refs.\cite{GRV98,MRSG,MRST2001,CTEQ6,Donnachie6}.}
\label{psiju10}
\end{figure}

To compare the two gluon exchange model with data on $J/\Psi$
photoproduction we need to specify the gluon parton distribution 
function. The PDF used in the following calculations 
are shown in Fig.\ref{psiju10} as a function of $x$ and for
$Q^2{=}m_J^2$. The upper axis of Fig.\ref{psiju10} indicates the
relevant invariant collision energy, $\sqrt{s}{=}m_J{/}\sqrt{x}$.
The old GRV98~\cite{GRV98} and MRSG~\cite{MRSG} functions show
a very steep dependence at small $x$, while the most modern
MRST2001~\cite{MRST2001} and CTEQ6~\cite{CTEQ6} functions are almost flat
in $x$. The DL~\cite{Donnachie6} structure function shows some
average dependence between the two extreme GRV98 and MRST2001
cases.

The PDF shown in Fig.\ref{psiju10} are very similar for
$x{\ge}0.01$, which corresponds to the invariant collision energy
of$\sqrt{s}{<}30$~GeV. That means that the data available 
before H1 and ZEUS measurement
are not sensitive to the choice of PDF. One might expect that the
differences between the different gluon functions in reproducing the 
$\gamma{+}p{\to}J/\Psi{+}p$ reaction should appear only at high 
energies. Since  $xg(x,Q_0^2)$ enters the photoproduction cross section
as an unintegrated function  squared, the sensitivity of 
the energy dependence of $J/\Psi$ photoproduction to the type of 
PDF should be extremely high. 

\begin{figure}[b]
\vspace*{-4mm}
\hspace*{-5mm}\psfig{file=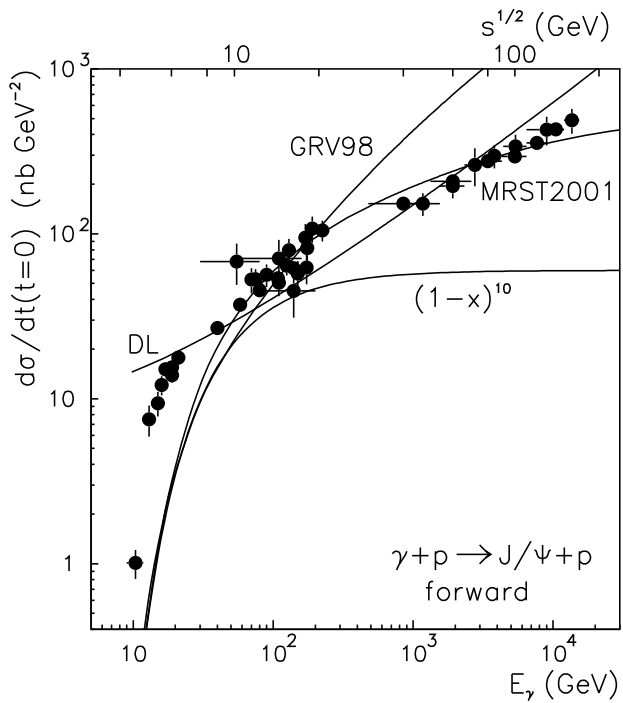,height=9.5cm,width=9.5cm}\vspace*{-4mm}
\caption[]{The forward $J/\Psi$ photoproduction cross 
section~\cite{Holmes,Adloff,Chekanov} as a function of photon 
energy, $E_\gamma$. The lines show the pQCD calculations by
two gluon exchange model with the gluon distribution functions 
GRV98~\cite{GRV98}, MRST2001~\cite{MRST2001} and DL~\cite{Donnachie6}.
The upper axis indicates the invariant collision energy.}
\label{psiju1}
\end{figure}

The calculations were done for two extreme cases given by
GRV98 and MRS2001 and for the DL function. The energy dependence
of the QCD coupling constant $\alpha_s$ was taken from Ref.~\cite{PDG}.
The mass of the charm quark allows for  additional
freedom in the absolute normalization of the calculations and we used
$m_q$=1.3~GeV. Fig.\ref{psiju1} shows the forward
$\gamma{+}p{\to}J/\Psi{+}p$ cross section as a function of photon 
energy in comparison to the two gluon exchange model calculations
by Eq.(\ref{twog}) with different gluon distribution functions.

As expected, the different PDF equally well reproduce the 
data around $\sqrt{s}{\simeq}10$~GeV. The GRV98 function is too steep
at small $x$ to describe the $J/\Psi$ photoproduction at high
energies. The DL function reasonably well reproduces the data over a
wide range of photon energies, but has incorrect threshold 
dependence, because of its asymptotic behaviour at large 
$x{\simeq}1$, as can be seen from Fig.\ref{psiju10}.

The modern MRST2001 gluon distribution function describes 
the energy dependence of the forward $J/\Psi$ photoproduction
cross section very well, providing as well the correct threshold behaviour.
It is also important to note that the two gluon model calculations with
MRST2001 leave some room for contribution from other
processes at $\sqrt{s}{<}$8~GeV. As is illustrated by 
Figs.\ref{psiju6},\ref{psiju7} and \ref{psiju2a} the data 
actually require a different photoproduction mechanism 
close to threshold, which certainly can be added to the two
gluon exchange contribution.

The $t$ dependence of the two gluon exchange is introduced 
artificially through the proton EM form factor $F_1$ 
providing the slope $b$=4.6~GeV$^{-2}$ at $t$=0. Although
this result is in a agreement with the data shown in
Fig.\ref{psiju2a} one may consider this agreement as 
a reasonable guess.
 
\section{Photon gluon fusion model.}
In the photon-gluon fusion 
model~\cite{Fritzsch,Jones,Leveille,Barger,Duke,Weiler,Roberts5}
(PGF) the photon fuses with a 
gluon from the nucleon and forms a $c{\bar c}$ pair, as is 
shown in diagram in Fig.\ref{psiju14}b). Since a gluon transforms 
as a color octet, the produced $c{\bar c}$ pair must radiate a soft gluon
before the hadronization into the color singlet final state.
Therefore the PGF presumably consider an inclusive charmonium or
open charm photoproduction associated with excitation of the
final nucleon.

The model convoluted the the gluon momentum distribution and the
photon-gluon cross section $\sigma_{\gamma{g}{\to}c{\bar c}}$,
which is given as 
\begin{eqnarray}
\sigma_{\gamma{g}{\to}c{\bar c}}{=}\frac{2\pi\alpha \, e_c^2 \alpha_s}
{{\tilde s}^3} ( [{\tilde s}^2+4m_c^2({\tilde s}^2-2m_c^2)]
\nonumber \\ \times\ln\left[\frac{1+\beta}{1-\beta}\right] 
-\beta[{\tilde s}^2+4{\tilde s}m_c^2]),
\end{eqnarray}
where $e_c$ and $m_c$ are the charge and mass of the charm quark, 
respectively, $\alpha_s$ is the QCD coupling constant,
$\alpha$ is electromagnetic coupling constant, ${\tilde s}$ is squared 
invariant mass of the photon-gluon system and
\begin{equation}
\beta^2 =1-\frac{4m_c^2}{{\tilde s}}.
\end{equation}

The PGF does not consider explicitly the formation of the final state,
which should enter through the $J/\Psi$ wave function as in
the case of two gluon exchange model. Instead the convolution of
photon-gluon cross section with the gluon momentum distribution 
is performed over the range from charmonium production threshold, 
$2m_c$, up to the open charm production threshold $2m_D$, given by the
mass of $D$ meson. Finally the charmonium photoproduction
cross section is
\begin{equation}
\sigma {=} f\!\int\limits_{4m_c^2}^{4m_D^2} \frac{d{\tilde s}}{{\tilde s}}
\, \, \sigma_{\gamma{g}{\to}c{\bar c}} \,\,
g(x{=}\frac{{\tilde s}}{s},Q_0^2),
\end{equation}
where the factor, $f$, is an adjustable parameter that accounts for the 
fraction of the specific charmonium  bound states available in 
the mass region between $2m_c$ and $2m_D$. For the $J/\Psi$ meson 
we fit it as $f$=0.062 with $m_c$=1.3~GeV. Furthermore, we
take $Q_0^2$=$m_J^2$.

\begin{figure}[b]
\vspace*{-4mm}
\hspace*{-5mm}\psfig{file=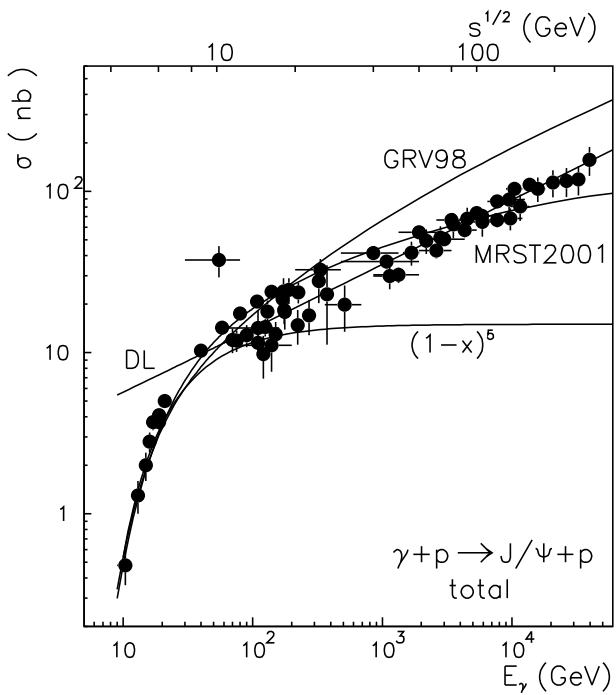,height=9.5cm,width=9.5cm}\vspace*{-4mm}
\caption[]{The total $J/\Psi$ photoproduction cross 
section~\cite{Holmes,Frabetti,Aid,Breitweg1,Adloff,Chekanov}
as a function of photon energy, $E_\gamma$. 
The lines show the pQCD calculations by
photon-gluon fusion model with the gluon distribution functions 
GRV98~\cite{GRV98}, MRST2001~\cite{MRST2001} and DL~\cite{Donnachie6}.
The upper axis indicates the invariant collision energy.}
\label{psiju3}
\end{figure}

The PGF calculations with different gluon distribution functions
are shown in Fig.\ref{psiju3}, together  with data on the total
$\gamma{+}p{\to}J/\Psi{+}p$ reaction cross section. Obviously, the
PGF is less sensitive to the type of gluon distribution functions
in comparison to the two gluon exchange model. The calculations with
MRST2001~\cite{MRST2001} and DL~\cite{Donnachie6} reproduce 
the data  better than the results obtained with GRV98~\cite{GRV98}, which
indicates a much steeper energy dependence. However the  MRST2001
and  GRV98 distributions both have correct threshold behaviour, while DL - not.
In principle, the DL gluon distribution function can be corrected 
in order to get the correct asymptotic behaviour as $x{\to}1$, by 
multiplying it with  $(1-x)^5$. However, then the agreement with 
the data disappears.

The PGF model does not contain a $t$ dependence and in the best 
case it can be artifically introduced through the proton EM
form factor. Although the PDF calculations are able to reproduce the 
cross section for
$J/\Psi$ photoproduction starting from threshold we would not claim 
that pQCD is able to describe the data at $\sqrt{s}{\le}$10~GeV.
We consider this good agreement between the data and PGF
calculations due to the correct asymptotic behaviour of the
gluon distribution function as $x{\to}1$, given by 
${\simeq}(1{-}x)^5$ dependence. Fig.\ref{psiju3} shows this function
alone in order to illustrate its threshold dependence. Furthermore,
in the two gluon exchange model the PDF enters  squared and 
the near threshold behaviour of the $J/\Psi$ photoproduction
cross section is then given by function ${\simeq}(1{-}x)^{10}$,
which is shown in Fig.\ref{psiju1}. However, now the dependence is
too steep to reproduce the data near the threshold.
 
\section{Conclusion.}
We have analyzed the available data on $J/\Psi$ photoproduction in terms of 
pomeron exchange, two gluon exchange and photon-gluon fusion models.

Allowing the pomeron-quark interaction to be flavour dependent and
introducing the soft and hard pomerons it is possible to reproduce
almost all available data on $J/\Psi$ photoproduction at
$\sqrt{s}{>}$10~GeV. At the same time, the model reproduce well
the data on $\omega$ meson photoproduction, again taking into account
that the interaction of the pomeron with light and charm quarks is 
very different. However the pomeron exchange model cannot not reproduce
the data at $\sqrt{s}{<}$10~GeV and differential spectra at
large $|t|$. The agreement between the calculations and total
$J/\Psi$ photoproduction cross section near  threshold should 
be considered as accidental, because both the slope $b$ of the 
$t$ dependence and the forward extrapolated cross section are in
strong systematic disagreement with the pomeron exchange model at
$\sqrt{s}{<}$10~GeV.

The two gluon exchange pQCD calculations indicate strong sensitivity 
to the gluon distribution function. The results obtained with the most
modern MRST2001~\cite{MRST2001} and DL~\cite{Donnachie6} PDF 
reproduce the forward $J/\Psi$ photoproduction cross section
at $\sqrt{s}{>}$10~GeV quite well. The calculations with MRST2001 show correct 
threshold dependence of the cross section and leave  room for 
a contribution from other mechanisms at low energies. The two gluon 
exchange model does not contain any $t$ dependence. Its artificial 
introduction through the isoscalar EM proton form factor allows one to 
reproduce the slope of the $t$ dependence at high energies, but fails at 
$\sqrt{s}{<}$10~GeV.

The calculations with the photon-gluon fusion model again show sensitivity
to the gluon distribution function. The PGF results calculated with
MRST2001~\cite{MRST2001} and DL~\cite{Donnachie6} PDF are in reasonable 
agreement with the data on the total $J/\Psi$ photoproduction cross section.
The PGF does not constrain the $t$ dependence.

Although the PGF calculations with MRST2001 reproduce the threshold behaviour 
of the total $\gamma{+}p{\to}J/\Psi{+}p$ cross section it does not
prove that pQCD is able to describe the reaction at $\sqrt{s}{<}$10~GeV.
It is just an artifact of the correct PDF asymptotic behaviour at
$x{\to}1$, given by ${\simeq}(1{-}x)^5$ dependence. For instance it is 
not a case for two gluon exchange model, where the PDF enters as squared 
and finally the ${\simeq}(1{-}x)^{10}$ dependence is too steep to reproduce
the $J/\Psi$ photoproduction data near the threshold. 

We allocate the $J/\Psi$ photoproduction at low energies and large 
$|t|$ to the mechanism different from pomeron or two gluon exchanges.
The energy dependence of the differential cross section, $d\sigma{/}dt$,
indicates that this mechanism does not depend on $s$.
Furthermore, small slope $b$ of the $t$ dependence indicates that
the exchange does not couple to the isoscalar EM proton form factor.
We consider that this might be axial vector  trajectory 
exchange that couples to the  axial form factor 
of the nucleon. However, a definite conclusion requires more 
detailed study.

\acknowledgments{
A.S would like to acknowledge the warm hospitality and partial support 
of the CSSM during his visit. This work was supported by  the grant 
N. 447AUS113/14/0 by the Deutsche Forschungsgemeinschaft and the 
Australian Research Council.}

\end{document}